\documentclass[twocolumn]{webofc}

\usepackage{caption}
\usepackage{makecell}
\usepackage{booktabs}
\usepackage{multirow}
\usepackage{tabularx}
\usepackage{listings}

\newcolumntype{L}[1]{>{\hsize=#1\hsize\raggedright\arraybackslash}X}
\newcolumntype{R}[1]{>{\hsize=#1\hsize\raggedleft\arraybackslash}X}
\newcolumntype{C}[1]{>{\hsize=#1\hsize\centering\arraybackslash}X}
\newcolumntype{E}{>{\raggedleft\arraybackslash}X}
\newcolumntype{M}{>{\centering\arraybackslash}X}

\newcommand{\ra}[1]{\renewcommand{\arraystretch}{#1}}

\usepackage{xcolor}

\definecolor{codegreen}{rgb}{0,0.6,0}
\definecolor{codegray}{rgb}{0.5,0.5,0.5}
\definecolor{codepurple}{rgb}{0.58,0,0.82}
\definecolor{backcolour}{rgb}{0.95,0.95,0.92}

\lstdefinestyle{codehl}{
  backgroundcolor=\color{backcolour}, commentstyle=\color{codegreen},
  keywordstyle=\color{magenta},
  numberstyle=\tiny\color{codegray},
  stringstyle=\color{codepurple},
  basicstyle=\ttfamily\footnotesize,
  breakatwhitespace=false,         
  breaklines=true,                 
  captionpos=b,                    
  keepspaces=true,                 
  numbers=left,                    
  numbersep=5pt,                  
  showspaces=false,                
  showstringspaces=false,
  showtabs=false,                  
  tabsize=2
}

\begin{document}

\title{On-Premise Artificial Intelligence as a Service for Small and Medium Size Setups}

\author{
    \firstname{Carolina} \lastname{Fortuna}
    \inst{1}
    \thanks{
        \email{carolina.fortuna@ijs.si}
    } 
    \and
    \firstname{Din} \lastname{Mu\v{s}i\'{c}}
    \inst{1,2}
    \thanks{
        \email{din.music@ijs.si}
    }
    \and
    \firstname{Gregor} \lastname{Cerar}
    \inst{1}
    \thanks{
        \email{gregor.cerar@ijs.si}
    }
    \and
    \firstname{Andrej} \lastname{\v{C}ampa}
    \inst{3}
    \thanks{
        \email{andrej.campa@comsensus.eu}
    }
    \and
    \firstname{Panagiotis} \lastname{Kapsalis}
    \inst{4}
    \thanks{
        \email{pkapsalis@epu.ntua.gr}
    }
    \and
    \firstname{Mihael} \lastname{Mohor\v{c}i\v{c}}
    \inst{1}
    \thanks{
        \email{miha.mohorcic@ijs.si}
    }
}

\institute{
    Jozef Stefan Institute, Jamova 39, Ljubljana, 1000, Slovenia.
    \and
    Jozef Stefan Institute Postgraduate School, Jamova 39, Ljubljana, 1000, Slovenia.
    \and
    Comsensus, Verov\v{s}kova ulica 64, Ljubljana, 1000, Slovenia.
    \and
    National Technical University of Athens, Iroon Polytechneiou 9, Attiki, 15773, Greece.
}

\abstract{
    Artificial Intelligence (AI) technologies are moving from customized deployments in specific domains towards generic solutions horizontally permeating vertical domains and industries. For instance, decisions on when to perform maintenance of roads or bridges or how to optimize public lighting in view of costs and safety in smart cities are increasingly informed by AI models.
    While various commercial solutions offer user friendly and easy to use AI as a Service (AIaaS), functionality-wise enabling the democratization of such ecosystems, open-source equivalent ecosystems are lagging behind.
    In this chapter, we discuss AIaaS functionality and corresponding technology stack and analyze possible realizations using open source user friendly technologies that are suitable for on-premise set-ups of small and medium sized users allowing full control over the data and technological platform without any third-party dependence or vendor lock-in. 
}

\maketitle

\section{Introduction}
\label{intro}

Artificial Intelligence (AI) technologies have been mostly used by experts to create new services and manage the extraction of value from large amount of data enabling the rise of some technology giants. However, recently they are moving from customized deployments in specific domains towards  more generic solutions aiming for horizontally permeating various vertical domains and industries~\cite{makridakis2017forthcoming}. For instance, decisions on when to perform maintenance of roads or bridges \cite{martinez2020comparative} or how to optimize public lighting \cite{de2016intelligent} in view of costs and safety in smart cities are increasingly informed by AI models. However, to be successful in increasing our overall productivity, the adoption by non-experts in their day-to-day use is further supported by efforts to lower the entry barrier by democratizing AI \cite{banifatemi2021democratizing} through the development of intuitive, easy to set up, manage and use systems such as AI as a Service (AIaaS).

\begin{figure}[hbt]
    \centering
    \includegraphics[width=\columnwidth, keepaspectratio]{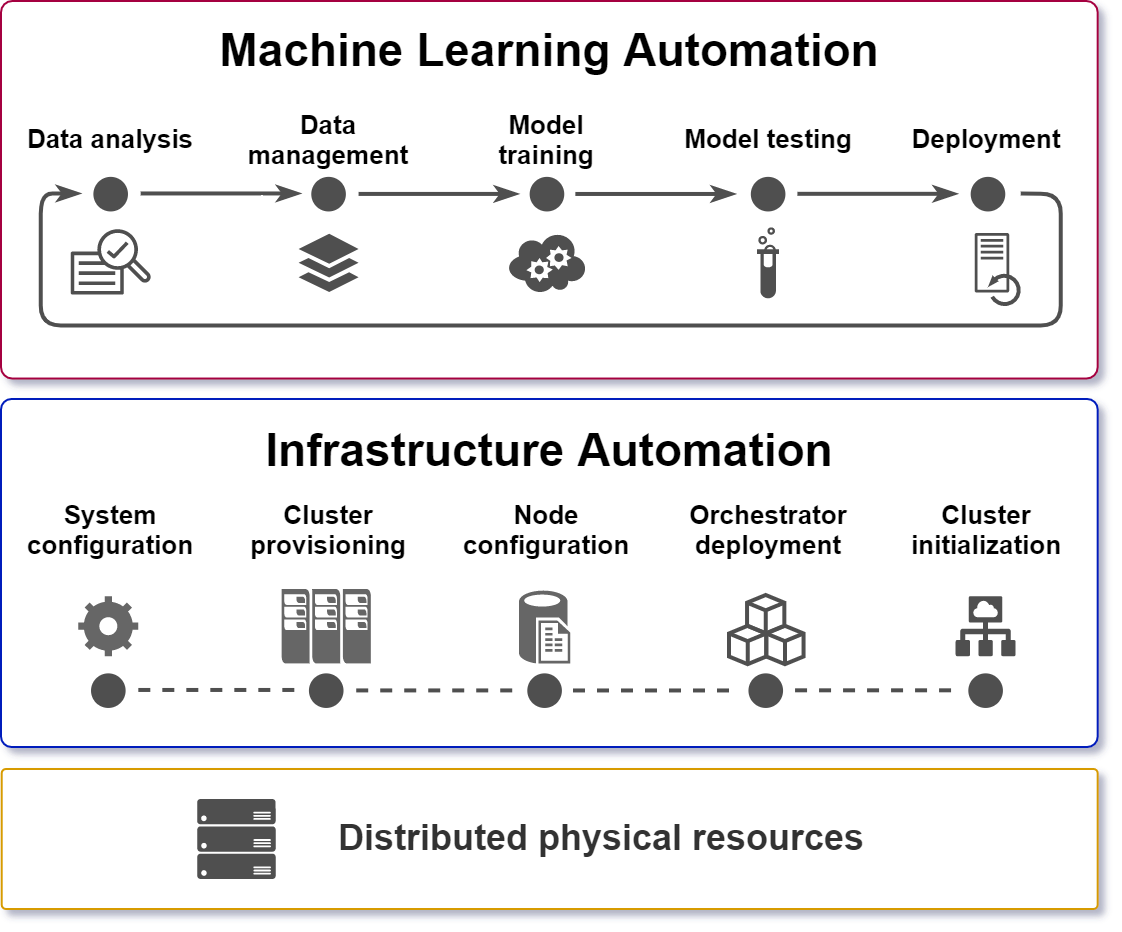}
    \caption{AIaaS technology stack.}
    \label{fig:aiaas}
\end{figure}

Figure \ref{fig:aiaas} presents a three layer abstraction of the AIaaS technology stack consisting of physical resources, infrastructure automation and machine learning automation, while an alternative view of a three layer abstraction focused less on infrastructure automation and more on AIaaS for different target groups is presented in  \cite{huang2018artificial}. Existing cloud service providers initially developed computing and storage functionalities such as Storage-, Infrastructure-, Platform- as a Service that were able to abstract physical resources and thus enabled an ecosystem for accelerated application development and scale-up of various companies \cite{goyal2014public}.  Subsequently, they added additional layers of abstraction and functionality leading to the so-called XaaS that enable ever more advanced application development and value creation \cite{hon2022cloud}.

Various commercial solutions already offer user friendly and easy to use AIaaS solutions, functionality-wise enabling the democratization of such systems, but their business and technological models rely on controlling parts of the AIaaS technology stack. While cloud providers mostly own the physical resources and keep them off the users' premises \cite{hon2022cloud}, recent commercial offerings are also enabling on-premise infrastructure deployment of so-called edge cloud computing \cite{9524558} where the physical resources may be on-premises while the infrastructure automation is managed by the provider. Fully controlled on-premises AIaaS stack is challenging and costly to deploy, and typically only available to large companies that have sufficient resources to realize it. 

In this chapter we discuss the infrastructure and machine learning automation for AIaaS stacks such as the one depicted in Figure \ref{fig:aiaas}. We discuss the functionality and corresponding technology stack for these two layers and analyze possible realizations using open source user friendly technologies that are suitable for on-premise set-ups of small and medium sized users allowing full control over the data and technological platform without any third-party dependence or vendor lock-in. 

Section \ref{sec:infra-auto} focuses on infrastructure automation, Section \ref{sec:ml-auto} focuses on machine learning automation while Section \ref{sec:conclusions} concludes the chapter.

\section{Infrastructure Automation}
\label{sec:infra-auto}

\begin{figure}[hbt]
    \includegraphics[width = \columnwidth, keepaspectratio]{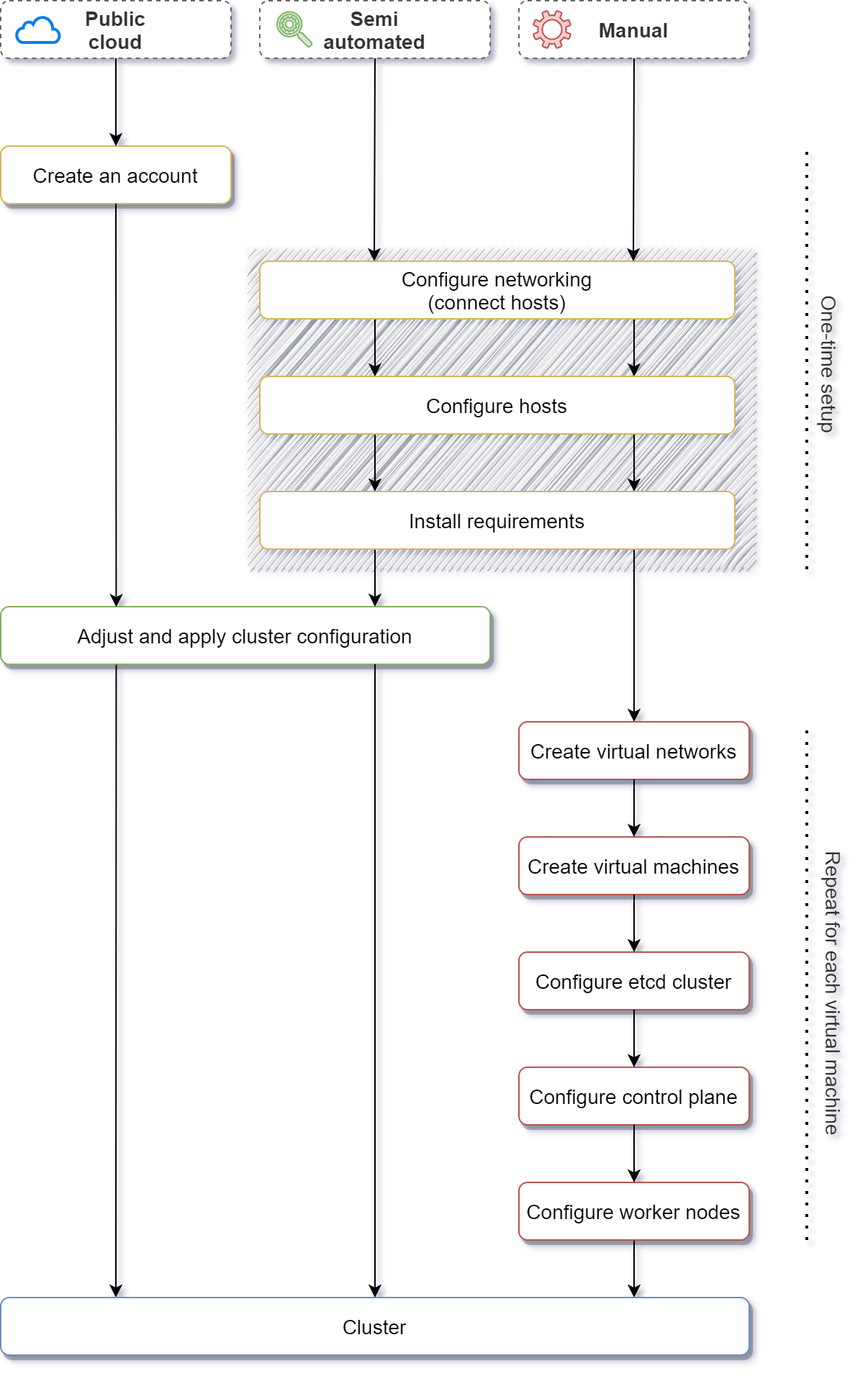}
    \caption{Comparison of infrastructure deployment steps on public cloud,  semi-automated environments and manual installation.}
    \label{cloud-tkk-manual}
\end{figure}

Cloud computing is already used in many businesses today \cite{MARSTON2011176} while many others are migrating to public clouds (i.e., AWS, GCP and MA) to reduce operational and management costs~\cite{bibi2012business}. The advantage of such solutions is the simplicity involved in managing the underlying infrastructure where few clicks or few line changes in configuration files reflect in production as depicted on the left in Figure~\ref{cloud-tkk-manual}. Alternatively, when public cloud solutions are discouraged for strategic, political or financial reasons, operational costs can also be reduced by adopting open source tools such as OpenStack, Apache Mesos and Eucalyptus or using these to replace proprietary infrastructure management tools where already in place. The aforementioned open source solutions are well suited for large-scale environments such as data centres, but add unnecessary complexity and overhead for most smaller clusters~\cite{corradi2014}.

For smaller, non-enterprise setups, such as hobby projects, home labs, or small and micro enterprises, it is hard to find a maintained solution that covers day to day needs. For such needs, one approach is to manually set up a cluster as depicted on the right side in Figure~\ref{cloud-tkk-manual}. Manual installation requires many steps to successfully set up a cluster. The first three steps that need to be performed only once are related to preparing the hosts by installing and configuring the hypervisors and possibly some other required software as marked with the grey background in Figure~\ref{cloud-tkk-manual}. In public clouds, this step is not required, as the virtualized layer is prepared in advance by the operator and later all users share the same infrastructure. Next follows a series of steps, marked with red under the \textit{Manual} process on the right in Figure~\ref{cloud-tkk-manual}, where each virtual machine must be manually created and configured.
This process is prone to human errors and can take days or even weeks, depending on the size of the cluster. In case the cluster needs to be upgraded, each node must be manually reconfigured, which takes almost as much time as setting up a new cluster. These steps can be automated, resulting in a so-called \textit{Semi-automated} process as illustrated in the middle in Figure~\ref{cloud-tkk-manual}.

Semi-automated solutions \cite{capuccini2019demand, libvirt-k8s-provider} differ from public clouds in that the hosts still need to be preconfigured, while the other steps are automated.
In fact, the host configuration process could also be automated by the end user, however, due to large differences between the environments, existing and emerging solutions assume manual configuration. 

Infrastructure automation lifecycle in its most general form has four independent phases: cluster creation, scaling, upgrading and destruction.

\subsubsection*{Creation phase}

\begin{figure}[hbt]
    \centering
    \includegraphics[width=\columnwidth, keepaspectratio]{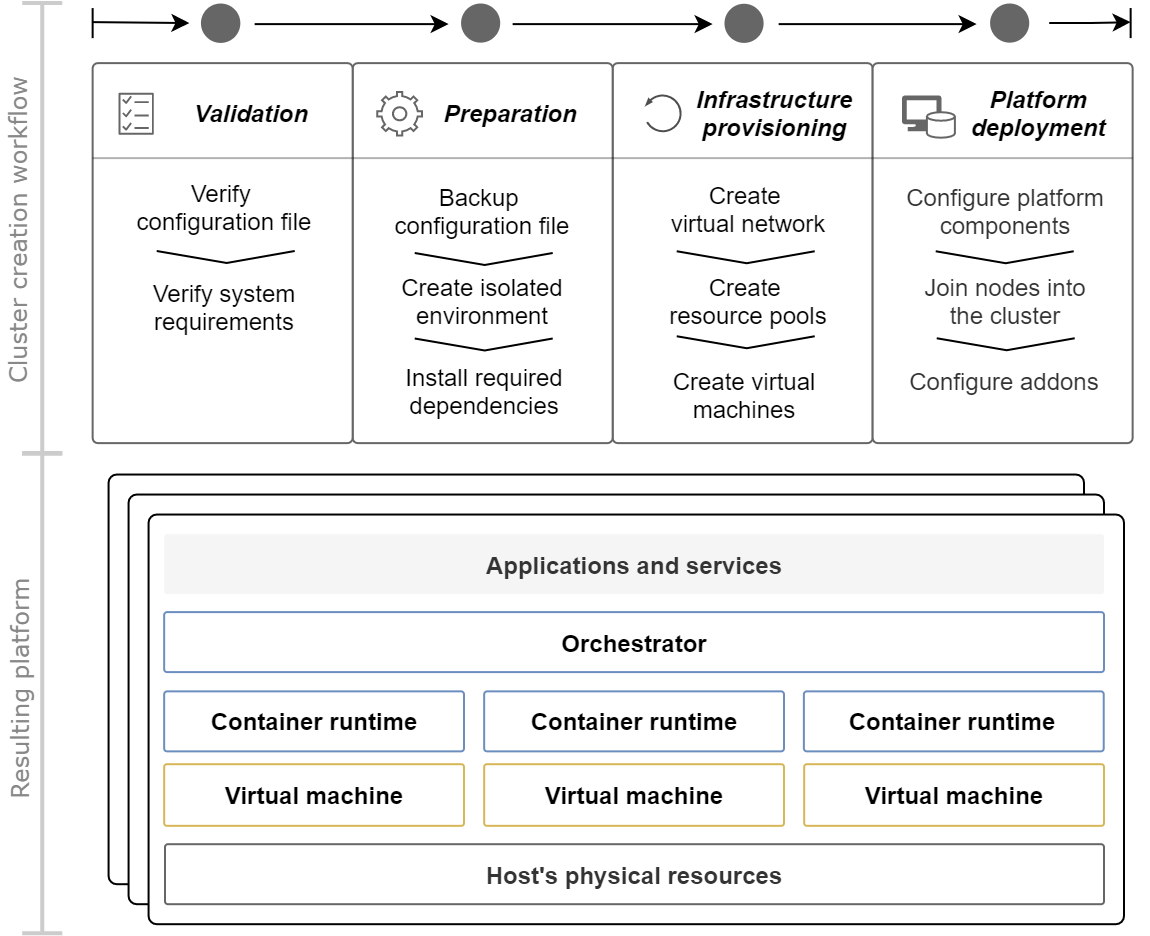}
    \caption{Infrastructure automation workflow and scheme of the resulting infrastructure.}
    \label{fig:mlops-infra-flow}
\end{figure}

As depicted in Figure \ref{fig:mlops-infra-flow}, cluster creation can be realized through the following four steps: validation, preparation, provisioning and deployment.

\paragraph*{Validation} 
The main objective of the validation step is to detect any potential errors as soon as possible.
Depending on the user input, which can be either a terminal command or a set of configuration files, the actions triggered by the input must be predicted and checked whether they can cause potential inconveniences later in the cluster creation process.
The validation process is critical to prevent unnecessary or potentially dangerous changes to the system and to save the user's time.
Validation of input is followed by validation of system requirements.
Based on user input, specific system requirements are verified. 

\paragraph*{Preparation} 
The preparation step ensures that missing packages, services, and dependencies are installed.
However, conditions that cannot be met during the preparation step must abort the cluster deployment.

\paragraph*{Infrastructure provisioning}
After the successful validation and preparation step, the infrastructure deployment step begins.
In this step, virtual components such as the virtual network, virtual machines and storage pools are provisioned according to the user's input.
On each virtual machine an operating system is deployed and configured to make it accessible on the local network.

At the end of this step, the provisioned infrastructure consists of independent virtual components that are not yet configured to be used as a cluster.
Therefore, an orchestrator needs to be deployed on top of the provisioned infrastructure.
The orchestrator simplifies the management and coordination of applications and services running in the cluster by treating the entire cluster as a single entity.

\begin{figure*}[hbt]
    \centering
    \includegraphics[width=0.7\linewidth, keepaspectratio]{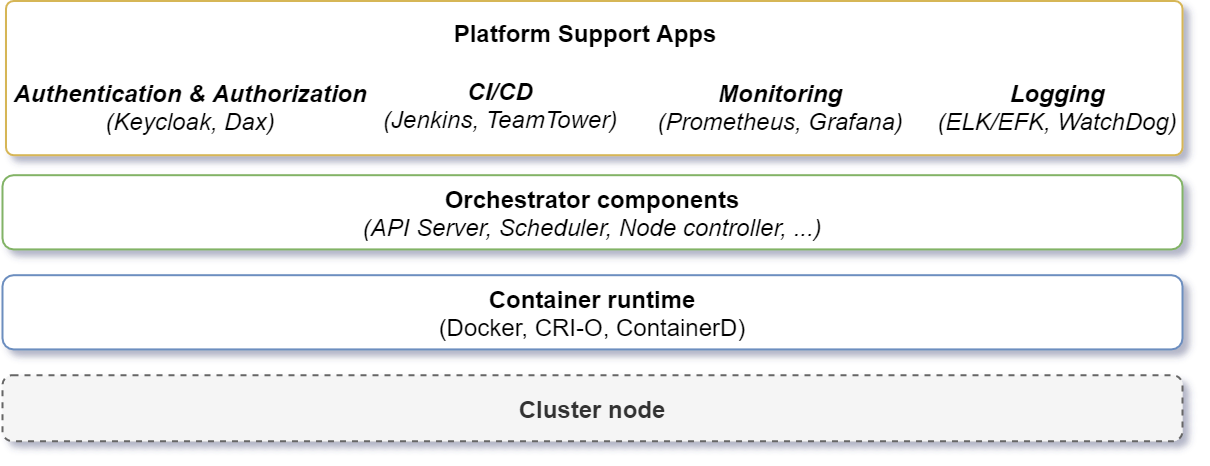}
    \caption{Example infrastructure resulting from the automation process.}
    \label{fig:infra}
\end{figure*}

\paragraph*{Platform deployment} 
Platform deployment step is responsible for the installation of orchestrator and platform components, and connection of all nodes into the cluster.
In this step, all orchestrator components, such as scheduler, server API, and other system controllers, are deployed on each provisioned virtual machine.
In addition to traditional virtualization, containerization can also be used.
Containerization is a popular, lightweight form of virtualization.
It takes advantage of the host's operating system kernel to create isolated processes in different system namespaces.
Compared to virtual machines, containers have low overhead because they do not require a dedicated operating system, making container creation, migration and deletion very efficient.
By combining containerization with traditional virtualization, virtual machines provide a high level of isolation, while containers reduce workload overhead and facilitate cluster management.
Containerization requires a container runtime, which can be installed and configured during this step.
After successful cluster installation, the cluster, as depicted in Figure \ref{fig:infra}, is ready for the deployment of custom workload.

In addition to the four general steps, infrastructure automation may also include other optional steps.
For instance, the installation of certain applications and services required on each deployed cluster, such as monitoring, logging, and authentication services, marked as \textit{Platform Support Applications} in Figure \ref{fig:infra}, can be performed as part of infrastructure automation.
However, the deployment of custom workloads, such as the machine learning operations pipeline, is typically separated from infrastructure automation to ensure loose coupling of the architecture.

\subsubsection*{Scaling, upgrading and destruction phases}

In addition to the initial setup of the cluster, the infrastructure automation strategy must also include the subsequent management of the cluster, such as scaling and upgrading the cluster.
As the workload increases or decreases, a running cluster may reach its capability limits or use its resources inefficiently. 
Therefore, scaling the cluster is an important part of the automation process.
Cluster scaling consists of adding and removing cluster nodes.
Scaling can be triggered either automatically based on collected cluster metrics or manually.
In larger environments consisting of many clusters, automatic scaling enables better overall utilization of resources.
Typically, resource usage of different clusters vary widely, and the ability to free up unused resources and make them available for other clusters to consume, reduces the required physical resources.
However, auto-scaling requires additional components to manage the infrastructure, which increases the complexity of the environment.
Therefore, for smaller environments that have more predictable resource consumption, manually triggered scaling may be sufficient.

The process of adding nodes to the cluster is similar to the initialization of the cluster.
The additional virtual machines are created and configured first and then the platform components are installed on them.
When the cluster nodes are ready, they are simply joined to the existing cluster.
Removing nodes, on the other hand, is more difficult because these nodes are already in operation and contain an active workload.
To gracefully remove a node, the node must first be drained, which means moving all of the workload to other running nodes.
This ensures that services remain available after the node is removed.

In addition to cluster scaling, the cluster must be regularly maintained to ensure that all components are up to date.
Compared to cluster scaling, upgrades still need to be automated, but are usually triggered manually.
This is because the compatibility of newer components must be checked in advance, otherwise the upgrade could render the entire cluster unusable.
There are many strategies for upgrading the cluster, but in-place upgrades are most common.
With in-place strategy, every node in the cluster is upgraded sequentially one-by-one.
This is done by draining the node, upgrading its components and rejoining the node back to cluster.
No additional virtual components need to be created for this approach, which is why it is frequently used, especially in smaller environments.

Once the cluster reaches the end of its life, it can be removed.
To remove the cluster completely, only the virtual components need to be stopped and deleted.
However, if the cluster contains valuable data, the data must be safely migrated before the cluster is destroyed, because it will be lost during this process.

\begin{figure*}[hbt]
    \centering
    \includegraphics[width=0.8\textwidth, keepaspectratio]{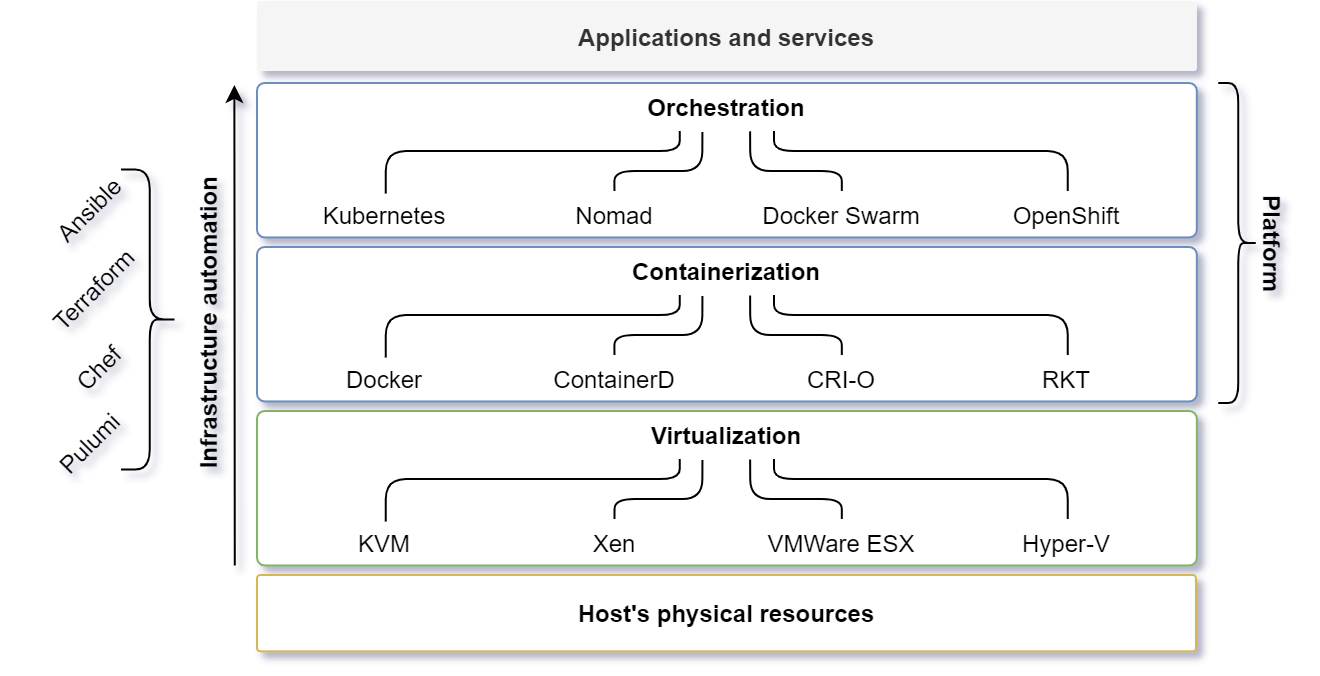}
    \caption{Infrastructure technology stack and popular automation tools.}
    \label{fig:mlops-on-prem-stack}
\end{figure*}

\subsection{Technology stack}
\label{sec:on-premise-stack}
Figure \ref{fig:mlops-on-prem-stack} shows the most popular open source technologies commonly used for infrastructure automation. A sensible combination of such technologies can be employed to realize an on-premises infrastructure according to the example depicted in Figure \ref{fig:infra}.

Technologies are divided into 4 groups comprising of Virtualization, Containerization, Orchestration and  Automation Tools.
Virtualization technologies represent hypervisors that are essential for traditional virtualization on top of physical resources.
They are capable of running multiple guest operating systems on a single physical host while isolating them from each other.
KVM and Xen are open source projects, with KVM being the most widely used.
Hyper-V and VMWare are good examples of proprietary alternatives.

The second group contains technologies for containerization.
Just as traditional virtualization requires a hypervisor to manage virtual resources, container runtimes enable containerization.
The most popular container runtimes are Docker, ContainerD, CRI-O, and RKT.

Container orchestrators can efficiently coordinate applications packaged as containers across multiple containerized environments.
The most popular container orchestrator is Kubernetes, which also supports most of the container runtimes mentioned above.
Nomad and OpenShift are seen as good alternatives for Kubernetes.
Docker Swarm, on the other hand, offers native support for orchestrating Docker environments.
However, it cannot be used with any other container runtime.

Finally, automation tools such as Ansible, Terraform, Pulumi, and Chef are used for provisioning a set of virtual machines, deploying container runtimes, and installing various dependencies.
All these Infrastructure-as-a-Code (IaaC) tools belong to the declarative languages, in which the desired state of the infrastructure is described rather than precisely defined with specific steps.
However, typical imperative languages such as Go and Python can be used to achieve the same result.
Imperative languages offer more freedom, but the automation process may take longer.

\begin{table*}[htbp]
    \caption{Comparison of semi-automated infrastructure automation solutions for small and medium set-ups.}
    \label{tab:semiauto-table}
    \ra{1.0}
    \centering
    \footnotesize
    \begin{tabularx}{\linewidth}{@{}L{0.5}C{0.7}C{0.7}C{0.7}C{0.7}C{0.7}}
        \toprule
            \textbf{\thead{ Projects\\\textbackslash{}\\Parameters}}
            & \textbf{\thead{Kubitect}}
            & \textbf{\thead{LKP}}
            & \textbf{\thead{KubeNow}}
            & \textbf{MicroK8s}
            & \textbf{K3s}
        \\\midrule
            On-premise traditional virtualization
            & Libvirt
            & Libvirt
            & OpenStack
            & /
            & /
        \\\midrule
            Management tool
            & CLI tool (kubitect)
            & Ansible scripts
            & CLI tool (kn)
            & CLI tool (microk8s)
            & Shell scripts
        \\\midrule
            Supported container runtimes
            & ContainerD
            & ContainerD, CRI-O, Docker
            & Docker
            & ContainerD
            & ContainerD
        \\\midrule
            Container orchestrator
            & Kubernetes
            & Kubernetes
            & Kubernetes
            & Kubernetes
            & Kubernetes
        \\\midrule
            Multi-host deployments
            & Y
            & N
            & Y
            & Y
            & Y
        \\\midrule
            Cluster scaling
            & Y
            & N
            & Y
            & Y
            & Y
        \\\midrule
            Cluster upgrades
            & Y
            & N
            & N
            & Y
            & Y
        \\\midrule
            High availability (HA) cluster topology
            & Y
            & N
            & N
            & Y
            & Y
        \\\midrule
            Out of the box storage solution
            & Rook
            & Rook
            & \textit{Provider dependant}
            & /
            & Longhorn
    \end{tabularx}
\end{table*}

\subsection{Semi-automated solutions for small and medium deployments}

In Table \ref{tab:semiauto-table} we have identified open source solutions that can be used to set up on-premise clusters. These solutions make use of the technologies discussed above and depicted in Figure \ref{fig:mlops-on-prem-stack}. 

All solutions analyzed in Table \ref{tab:semiauto-table}, Kubitect, Libvirt-k8s-provisioner (LKP), KubeNow, MicroK8s, and K3s, are classified as semi-automated because they have some unique requirements that must be met by the user.
Kubitect and LKP are capable of creating virtualized resources as part of the automation process. MicroK8s and K3s can also be used on top of the traditional virtualization, however, they do not provide infrastructure automation as part of the solution. Thus, without additional automation, MicroK8s and K3s nodes must be manually configured on each host.
KubeNow differs from other solutions as it can only be used with OpenStack, an open source private cloud solution, for on-premises deployments.

Kubitect, KubeNow and MicroK8s clusters are created and managed via custom command line interface (CLI) tools of the corresponding solutions.
K3s and LKP clusters, on the other hand, are managed directly via scripts, which are less intuitive to use.

All solutions use Kubernetes as an application and service orchestrator and, with the exception of KubeNow, primarily use ContainerD as a container runtime environment.
While KubeNow only supports Docker, LKP can be configured to use either Docker, CRI-O or ContainerD.
However, Docker and ContainerD are de facto standard and should cover most users' needs.

In terms of implementation, each solution uses a different set of technologies to create Kubernetes clusters.
The Kubitect CLI tool is written in Go and further takes advantage of Terraform and Ansible for infrastructure deployment and configuration.
LKP uses only Terraform and Ansible. Consequently, different scripts must be run to create or destroy the cluster.
MicroK8s and K3s, on the other hand, do not use Ansible or Terraform, but they also do not provide automation of the virtualized infrastructure.
Instead, the entire solution is implemented using a single programming language. 
MikroK8s is written in Python and K3s in Go.

Besides the implementation details, the solutions differ mainly in the supported topologies and management features.
LKP does not support cluster deployments across multiple physical hosts, which means that it is not possible to create truly high-availability (HA) clusters.
Likewise, it does not provide any other management functions, such as scaling and upgrading existing clusters.
Kubitect and KubeNow, on the other hand, support cluster deployment across multiple hosts.
However, only Kubitect is capable of creating HA clusters, as KubeNow is limited to a single control plane node whose failure renders the entire cluster unusable.
MicroK8s and K3s also support cluster deployments that span multiple hosts. However, since they do not have infrastructure automation out of the box, this must be done manually.

Every created cluster also requires a place to store the application's data.
While most storage solutions are deployed on top of the Kubernetes orchestrator and are able to consume disks attached to the host, some solutions provide configuration of storage clusters out of the box.
Kubitect and LKP both provide optional setup of distributed storage cluster that uses Rook as the storage orchestrator.
Using Rook, different kinds of storage types can be used, such as filesystem, object store and block storage, which covers all requirements of Machine Learning Operations (MLOps).
MicroK8s does not provide a distributed storage solution out of the box, neither as part of the cluster deployment nor as an optional add-on.
In comparison, K3s offers the Longhorn deployment as an optional add-on module that is equivalent to Rook in terms of features.
KubeNow is able to use the storage provided by the underlying cloud.

\begin{figure*}[htbp]
    \centering
    \includegraphics[width=0.9\linewidth]{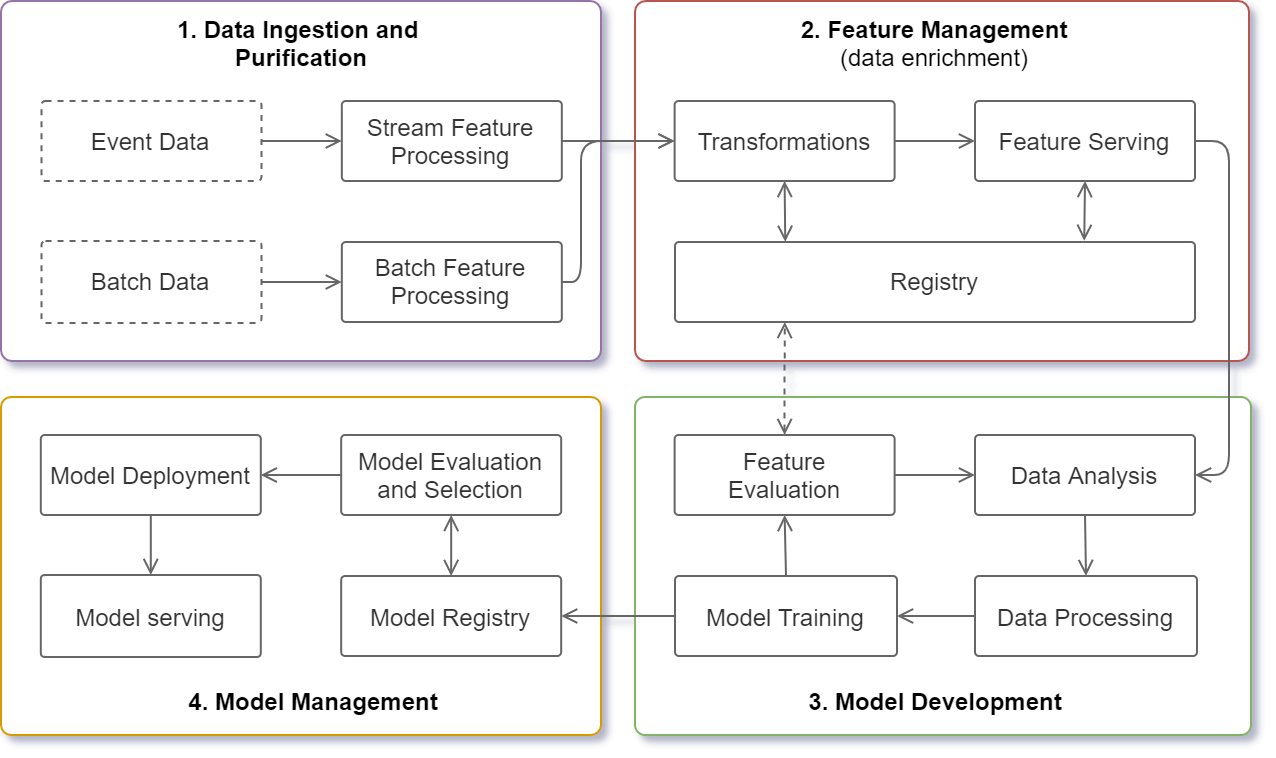}
    \caption{Machine learning automation workflow.}
    \label{fig:mlflow}
\end{figure*}
\section{Machine Learning Automation}
\label{sec:ml-auto}

The development of machine-learning models follows a well defined knowledge discovery process (KDP)~\citep{fayyad1996knowledge}. The main steps of the KDP consist of (1) data analysis, (2) data preparation (pre-processing), (3) model training and evaluation, and (4) model deployment~\citep{ruf2021demystifying}, as also represented in Figure~\ref{fig:mlflow}. In the past, such process and the enabling tools were familiar only to a limited number of domain experts and the process involved intense manual effort. However, in recent years, coordinated efforts have been taken by the private and public sectors to democratize AI and model development~\citep{allen2019democratizing} to empower less specialized users. 

The democratization process involves a division of labour and automation approach applied to the KDP, as elaborated more in details in~\citep{ruf2021demystifying}, where rather than a domain expert executing step-by-step the process in Figure~\ref{fig:mlflow} from the start to the end, the process only needs to be controled at a few key steps. Furthermore, generic models can also be trained and made available through an Inference as a Service (IaaS)~\cite{huang2018artificial} model where the users only leverage an existing, pre-trained model to make an inference used by their application or service. Such automation is enabled by MLOps~\citep{ruf2021demystifying}.

As can be seen from MLOps automation process depicted in Figure~\ref{fig:mlflow}, in the Data Ingestion and Purification phase, data from sensors or applications can be received in batch format such as hourly, daily or weekly batches or in a streaming format as it is being produced. The incoming data is then managed by feature stores rather than manually through generation scripts and file/database storage. Feature stores are data management services that harmonize data processing steps in producing features for different ML pipelines, making it more cost-effective and scalable compared to traditional approaches~\citep{patel2020unification}, and they can also be seen and made available as a service (i.e., STaaS). 

During model training, various combinations of features available in the feature store are used to train ML models. The model training phase can be made available to ML developers as MLaaS, where they can tune and customize desired models on desired features \citep{huang2018artificial}, or they can be entirely abstracted and hidden from less advanced users. The resulting models are stored in a registry, evaluated and some of them eventually deployed as part of the Model Serving phase. The model evaluation and deployment enable the development of the IaaS model where less experienced users are able to select and use pre-trained models for their applications.

\begin{figure*}[hbt]
    \centering
    \includegraphics[width=\linewidth, keepaspectratio]{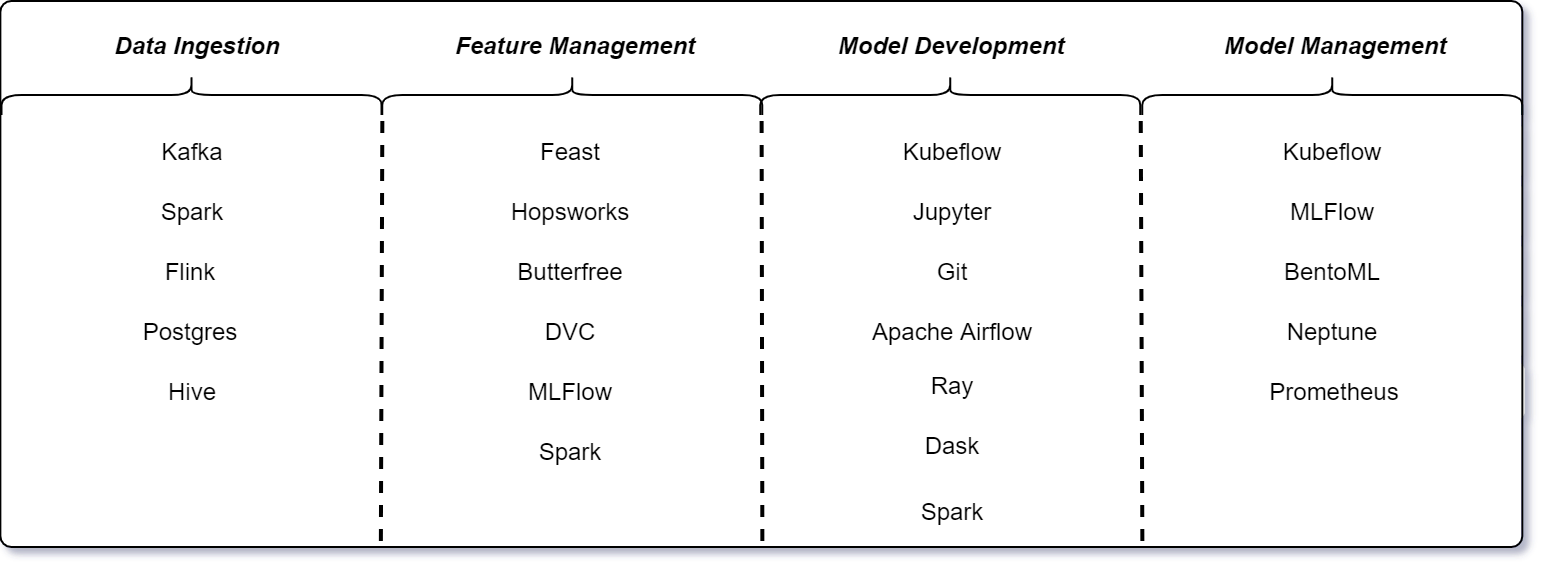}
    \caption{MLOps stack for on-premises deployment.}
    \label{fig:mlops-flow}
\end{figure*}

\begin{table*}[htbp]
\caption{List of open-source feature store solutions suitable for time series data.}
\label{tab:available-feature-stores}
\ra{1.3}
\centering
\footnotesize
\begin{tabularx}{\linewidth}{@{}L{0.7}C{0.4}L{1.4}L{1.4}L{1.4}L{}L{}}
    \toprule
    \textbf{\thead{Name}}
    & \textbf{\thead{Open\\Source}}
    & \textbf{\thead{Data Sources}}
    & \textbf{\thead{Offline Storage}}
    & \textbf{\thead{Online Storage}}
    & \textbf{\thead{\\Deployment}}
   
    \\\midrule
    
    Feast
    & Y
    & BigQuery, Hive, Kafka, Parquet, Postgres, Redshift, Snowflake, Spark, Synapse
    & BigQuery, Hive, Pandas, Postgres, Redshift, Snowflake, Spark, Synapse, Trino, custom
    & DynamoDB, Datastore, Redis, Azure Cache for Redis, Postgres, SQLite, custom
    & AWS Lambda, Kubernetes, local
    \\\midrule
    
    Hopsworks
    & Y
    & Flink, Spark, custom Python, Java, or Scala connectors
    & Azure Data Lake Storage, HopsFS, any SQL with JDBC, Redshift, S3, Snowflake
    & any SQL with JDBC, Snowflake
    & AWS, Azure, Google Cloud, local
    \\\midrule
    
    Butterfree
    & Y
    & Kafka, S3, Spark
    & S3, Spark Metastore
    & Cassandra
    & local
    \\\midrule
\end{tabularx}
\end{table*}

\subsection{Technology stack}
The choice of technologies available for realizing the ML automation workflow depicted in Figure~\ref{fig:mlflow} is becoming richer every year. In Figure~\ref{fig:mlops-flow}, we provide an overview of possible candidates. During the data ingestion phase, a batch can for instance be retrieved from a database or STaaS service (e.g., BigQuery, Hive, Postgres) or a stream processing platform (e.g., Kafka, Spark, Flink).

For managing the features, fully automated solutions such as feature stores (e.g., Feast, Hopsworks) can be employed or it can also be realized using data versioning and data processing solutions such as DVC, Pandas, and Spark. Several closed source feature stores have been recently developed, while to the best of our knowledge, three feature management solutions are available as open source with their characteristics summarised in  Table~\ref{tab:available-feature-stores}. It can be seen from the third column of the table that they include connectors to support fast interconnection with various storage solutions (e.g., BigQuery, S3, Postgres, ...) and streaming platforms (e.g., Kafka, Spark). As per columns three and four, it can be seen that all open source stores support offline and online storage such as public cloud provider's BigQuery, Azure, S3 and Snowflake or open source solutions such as PostgreSQL and Cassandra.  As can be seen from the sixth column of the table, the open source feature stores can be deployed locally and also in the public cloud.

For the model development phase, the necessary tools include tools that enable code development and versioning (e.g., Kubeflow, Jupyter and Git), tools that are responsible to orchestrate model training such as Kubeflow Pipelines, Airflow and CML, as well as tools for speeding up training through parallelization such as Ray, Dask and Spark. Using the set of tools in this phase, MLaaS can be realized where developers are presented with a complete playground for custom model development.

The model serving comprises tools for model tracking, evaluation and serving such as Kubeflow and MLFlow. Especially these two tools are evolving rapidly with largely overlapping functionality in the model development and service phases of MLOps. BentoML, MLEM, TensorFlow Serving (TFX) and Kale can also be used for serving, \textit{Neptune.ai} or \textit{``Weights \& Biases''} for versioning while Prometheus for model performance monitoring.

\begin{figure*}[htbp]
    \centering
    \includegraphics[width=0.7\linewidth]{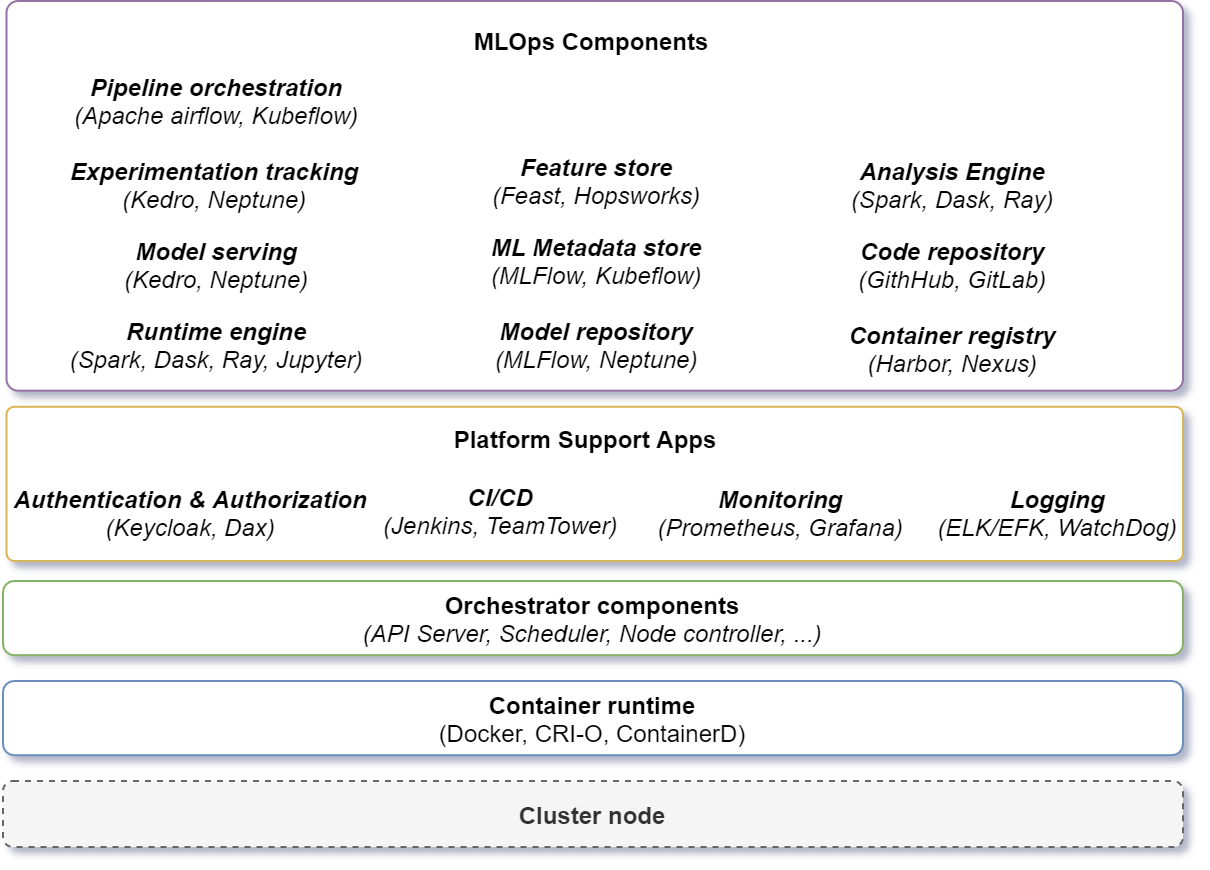}
    \caption{Example on-premise infrastructure for AIaaS.}
    \label{fig:ml-uc}
\end{figure*}

\subsection{On-premise automation use case}
Figure~\ref{fig:ml-uc} shows an example of an on-premise ML automation platform. To build a model, we first require data. A feature store (e.g., Feast) can be set up to help dealing with live/streaming sensor data (i.e., events) from stream processor (e.g., Kafka) and historical data from a database(s)/file(s) (i.e., data warehouse/lake). A registry within a feature store contains instructions for obtaining or composing features. The example in Listing~\ref{code:feast}, for instance,  defines mean hourly energy consumption as a feature. With all the instructions set and all data sources available, the feature store manages access to the features for model training and models deployed in production. In addition, a feature store unifies access to features for a development and production environment. Once a new instruction is added to the feature store's registry, a new feature becomes immediately available.

The model training phase depicted in Figure~\ref{fig:mlflow} focuses on analyzing data and building ML models with ML pipelines. Data analysis can lead to a discovery of new features that can be valuable addition for future models. Therefore, they are contributed to the feature store (in previous phase) as instructions in the registry to benefit anyone using it. Machine learning pipelines, which define process from data to machine learning model, are specified in code (e.g., Python script, Jupyter Notebook) with help from ML frameworks. Data for pipelines are obtained preferably from a feature store or other external sources. Developed ML models are pushed automatically or manually to the MLFlow service (see Fig.~\ref{fig:mlflow}).

The ML models can be developed using several open-source machine learning frameworks such as NumPy, SciPy, scikit-learn, XGBoost, PyTorch, TensorFlow, Keras, and JAX. In addition, these libraries can be extended with higher-order ``AutoML'' frameworks, such as AutoKeras or AutoSklearn. These frameworks automate search for optimal ML algorithms, architectures, and hyper-parameters for given data.

Tuning ML pipeline and therefore ML models is a complex process that usually requires multiple iterations. It should be noted that the feature importance in MLOps pipelines is assessed by the Model Training component depicted in Figure~\ref{fig:mlflow} subject to the availability of the features. 

\begin{lstlisting}[language=Python, caption={Example recipe for generating the average household energy consumption over 1 hour feature (energy\_mean).}, label={code:feast}]
residential_hourly_stats = FileSource(
  path=str(residential_dataset_path),
  event_timestamp_column='timestamp',
)

consumption_hourly_stats_view = FeatureView(
  name='residential_hourly_stats',
  entities=['residential_id'],
  # Measurement validity period. 
  # 1h in our case. Used when data is joined
  ttl=Duration(seconds=3600),
  # Describe features in the data
  features=[
    Feature(name='ts', dtype=ValueType.FLOAT),
    Feature(name='energy', dtype=ValueType.FLOAT),
  ],
  online=True,
  # source of the data (parquet file in our case)
  batch_source=residential_hourly_stats,
  tags={},
)
\end{lstlisting}

To make the best out of an ML model development process, we consider good practice to keep track of ML pipeline source code with versioning tools (e.g., Git) and have it co-located on centralized services (e.g., GitHub, GitLab). This way, the MLOps system (especially the orchestrator) knows where to find and access the latest source code. With a source code branching feature of versioning tools, we can develop features or improvements in parallel and test them independently before they might get merged into the main branch.

ML pipeline steps (including the final ML algorithm) can be time-consuming in many cases. It would be inefficient to rebuild ML models with the same data and parameters more than once. To minimize this overhead, a developed model is pushed into model store (e.g., MLFlow, KubeFlow) service for later retrieval. The MLFlow service in our case keeps track of ML model artifacts. The artifacts are, for instance, model hyper-parameters, self-evaluation metrics, model's score/grade, and model’s binary representation (i.e., inner state, model parameters, model weights) for later retrieval without (re)training. To keep the pipeline source code and artifacts connected for the record, MLFlow also preserves essential parts of source code, and source code commit hash. Code sample in Listing~\ref{code:mlflow} shows required changes to pipeline code to utilize MLFlow service.

\begin{lstlisting}[language=Python, caption={Example model development tracking with MLFlow.}, label={code:mlflow}]
# Start recording the run
with mlflow.start_run():
    model.fit(X_train, y_train)

    # predict values for evaluation
    y_pred = model.predict(X_test)

    # MLFlow will store model into pickle for us.
    mlflow.sklearn.log_model(
        sk_model=model,
        artifact_path='model',
        registered_model_name=config.REGISTERED_MODEL_NAME, # model registration here or manually on web UI
        pip_requirements=['-r ./requirements.txt'],
    )

    # Log all relevant metrics for given task
    rmse = metrics.mean_squared_error(y_test, y_pred, squared=False)
    mlflow.log_metric('RMSE', rmse)

\end{lstlisting}

Once models are preserved in model store (i.e., MLFlow), we can access and inspect them through the web or application programming interface (API). We can inspect individual models, their meta-data, input parameters and evaluation metrics, and compare them with other models. From a list of proposed/preserved models, we can manually (via web interface) or automatically (with software) promote or demote models into \textit{staging} (or testing) and \textit{production} grade, which would happen after selected models undergo additional inspection and evaluation steps. The assigned labels help test deployment and production deployment infrastructure to pick the correct model.

Once the deployment process is triggered, the automated pipeline takes labeled models (\textit{staging} or \textit{production}), converts them into containers suitable for the model serving framework (e.g., BentoML), and pushes ready-made containers into blob storage (e.g., RedisDB, MinIO). From blob storage, models can be quickly (re)deployed by model serving framework service. Once deployed, MLOps pipeline exposes their interaction interfaces to the world.

For orchestration (including automation and synchronization) of the MLOps pipeline, we use Apache Airflow. To orchestrate, Apache Airflow requires instructions made out of small tasks. Tasks can depend on each other, but their inter-dependencies must form a directed acyclic graph (DAG). Airflow can be used to trigger model rebuilding, preparing staging and production containers and (re)deploying models.

Our on-premise use case consists of many complex interconnecting services. For instance, Apache Airflow orchestrator, Feast feature store, JupyterHub, MLFlow model storage and RedisDB blob storage. However, most of these services consist of many smaller hidden building blocks. For example, MLFlow model storage requires an MLFlow server, a PostgreSQL database for storing models' metadata, and MinIO for storing models' artifacts. To reduce the complexity, the usual practice is to containerize individual building blocks and present service as a single (unsplittable) entity such as pod (in Kubernetes context) or compose (in Docker context).

\section{Conclusions}
\label{sec:conclusions}

In this chapter, we discussed the importance of AIaaS along with recent developments in automation that enable on-premises AIaaS deployments, thus extending the benefits of such systems also to small and medium size setups requiring full control over the data and technological platform. We first focused on the general process required for on-premise infrastructure deployment and identified a number of existing open source tools and technologies for possible realizations. Then we discussed the general process required for machine learning pipeline automation in view of enabling AIaaS and also proceeded at identification of suitable technologies for its on-premise implementation. Overall, we argued that the available open source technologies are sufficiently mature and suitable for small to medium size on-premise setups of AIaaS functionality, and can be automatically configured and interconnected in such a way to support easy and intuitive setting up, use and management, hiding the complexity from the non-expert users. Such AIaaS setups can thus support gradual introduction of smart services also to various stakeholders and organizational entities in cities and towns of various sizes, and enable their extension and scaling with the increasing needs and introduction of new services and applications.

\section*{Acknowledgments}
This work was funded in part by the Slovenian Research Agency under the grant P2-0016 and the European Commission under grant agreements 101000158 (MATRYCS) and 872525 (BD4OPEM).

\bibliography{bibliography}

\end{document}